%% file: main.tex
\documentclass[sigconf,natbib=true,square,numbers]{acmart}
\usepackage{array}
\usepackage{xspace}
\usepackage{listings}
\usepackage{xcolor}
\usepackage{appendix}
\usepackage{booktabs}
\usepackage{amsmath}
\usepackage{url}

\lstdefinestyle{promptstyle}{
    backgroundcolor=\color{gray!10},
    basicstyle=\ttfamily\fontsize{7}{9}\selectfont,
    frame=single,
    breaklines=true
}

\renewcommand\footnotetextcopyrightpermission[1]{}

\input{macros}
\AtBeginDocument{%
  \providecommand\BibTeX{{%
    \normalfont B\kern-0.5em{\scshape i\kern-0.25em b}\kern-0.8em\TeX}}}

\makeatletter
\gdef\@copyrightpermission{
 \begin{minipage}{0.3\columnwidth}
 \href{https://creativecommons.org/licenses/by/4.0/}{\includegraphics[width=0.90\textwidth]{figs/4ACM-CC-by-88x31.eps}}
 \end{minipage}\hfill
 \begin{minipage}{0.7\columnwidth}
 \href{https://creativecommons.org/licenses/by/4.0/}{This work is licensed under a Creative Commons 
Attribution International 4.0 License.}
 \end{minipage}
 \vspace{5pt}
}
\makeatother

\begin{document}

%\ok{Just a try, please revise}
% \title{Prompt-Based Document Modifications In Ranking Competitions}
\title{White Hat Search Engine Optimization using Large Language Models}
\settopmatter{printacmref=false, printfolios=false}


\author{Niv Bardas}
\email{niv.b@campus.technion.ac.il}
\affiliation{%
  \institution{Technion}
  \city{Haifa}
  \country{Israel}
}

\author{Tommy Mordo}
\email{tommymordo@technion.ac.il}
\affiliation{
  \institution{Technion}
  \city{Haifa}
  \country{Israel}
}

\author{Oren Kurland}
\email{kurland@technion.ac.il}
\affiliation{%
  \institution{Technion}
  \city{Haifa}
  \country{Israel}
}

\author{Moshe Tennenholtz}
\email{moshet@technion.ac.il}
\affiliation{%
  \institution{Technion}
  \city{Haifa}
  \country{Israel}
}

\author{Gal Zur}
\email{gal.zur@campus.technion.ac.il}
\affiliation{%
  \institution{Technion}
  \city{Haifa}
  \country{Israel}
}

\renewcommand{\shortauthors}{Bardas et al/.}

\input{abstract}

\maketitle

\input{intro}

\input{bots}
\input{experimental-setup}

\input{experimental-results}

\input{conclusions}

\input{appendix}

\balance
\bibliographystyle{ACM-Reference-Format}
% \bibliography{retrieval-methods}
\input{main.bbl}

\end{document}

%% file: macros.tex
\newcommand{\omt}[1]{}
\newcommand{\definedas}{\stackrel{def}{=}}
\newcommand{\firstmention}[1]{{\bf #1}}

\newcommand{\myparagraph}[1]{\vspace{0.3\baselineskip}\noindent{\textbf{#1}}.~}

\newcommand{\bt}{LLM agent}

\newcommand{\dc}{d_{curr}}

\newcommand{\dn}{d_{mod}}
\newcommand{\dni}{d_{mod}^{i}}

\newcommand{\trueteacher}{TT}

\newcommand{\firstDataset}{LambdaMARTComp\xspace}
\newcommand{\secondDataset}{E5Comp\xspace}
\newcommand{\sentReplace}{SentReplace\xspace}

\newcommand{\normFaith}{OrigFaith\xspace}
\newcommand{\topRet}{T}

\newcommand{\corpFaithE}{CorpFaith(E5)\xspace}
\newcommand{\normCorpFaithE}{CorpFaith(E5)\xspace}
\newcommand{\corpFaithT}{CorpFaith(TF.IDF)\xspace}
\newcommand{\normCorpFaithT}{CorpFaith(TF.IDF)\xspace}

\newcommand{\studDiff}{s}
\newcommand{\baseDiff}{r}

\newcommand{\statDif}{b}

%% file: abstract.tex
\begin{abstract}
% We present a set of prompting approaches for large language models to perform content-based white-hat search engine optimization; namely, legitimate document editing intended to improve the ranking of the document. An empirical study demonstrates the effectiveness of our methods.
We present novel white-hat search engine optimization techniques based on genAI and demonstrate their empirical merits.
\end{abstract}

%% file: intro.tex
\section{Introduction}
\label{sec:intro}
In competitive search settings \cite{Kurland+Tennnholtz:22a}, document
authors might respond to rankings induced for queries of interest by
modifying their documents. The goal is to improve
future ranking. These modifications are often referred to as search
engine optimization (SEO) \cite{Gyongyi+Molina:05a}. Our focus, as
that of prior work on competitive search
\cite{raifer2017information,goren2020ranking,Kurland+Tennnholtz:22a}, is on white hat SEO \cite{Gyongyi+Molina:05a}:
legitimate document modifications which do not hurt document quality
nor the search ecosystem.

We present novel document modification methods for ad hoc retrieval that are based on prompting large language models (LLMs). The desiderata of the modifications are (cf.,
\cite{goren2020ranking}): (i) potentially improving future ranking for a query at hand, (ii)
maintaining ``faithfulness'' to the original document, and (iii) producing a document of high (content) quality. Some of our methods are inspired by recent work on using
prompting techniques to have LLMs induce rankings in response to
queries \cite{Liang+al:22a,qin2023large,ma2023zero}. For example, some of the
prompts include information induced from past rankings to provide the
LLM with implicit information about the undisclosed ranking function.

We evaluate the document modification methods using datasets that
resulted from past ranking competitions
\cite{raifer2017information,Mordo+al:25a}. In addition, we organized
ranking competitions and used our modification methods as bots
competing against humans\footnote{The dataset of the competition we
  organized, and the accompanying code, will be made public upon
  publication of this paper. They are available for reviewing purposes
  at \url{https://github.com/promptdriven2025}.}. The empirical
evaluation demonstrates the merits of the most effective methods we
studied with respect to humans and to a highly effective feature-based
approach for document modification
\cite{goren2020ranking}.

  \myparagraph{Related work} We refer the reader to \cite{Kurland+Tennnholtz:22a} for a survey on competitive search. The work most related to ours is that of
  Goren et al. \cite{goren2020ranking} who devised a supervised
  feature-based document modification method. We discuss the method in Section \ref{sec:experimental-setup} and use it as a baseline. 

There is recent work on performing SEO for Web pages by prompting LLMs
\cite{Aggarwal+al:24a}. The prompts are designed for commercial search
engines such as Google and are not well suited to the
general ranking functions we use for evaluation. In addition, the
prompts, in contrast to ours, do not include examples of past rankings
and are mainly zero shot.
  
There is work on attacks (not necessarily white hat) on
BERT-based document rankers (e.g., \cite{Wu+al:23a}) and on LLM-based document and product rankers, conversational agents and question answering systems
(e.g., \cite{Kumar+Lakkaraju:24a,Pfrommer+al:24a,Nestaas+al:24a}). Our focus, in contrast, is on white hat content modification for document ranking which is not committed to a specific ranking function. We focus on dense and sparse retrieval in our evaluation.

\omt{
  The method selects a
sentence from the document and replaces it with a sentence from a
document which was the highest ranked for the query in the past. The
features used by the method include estimates for query term occurrence (to improve ranking) and estimates of the semantic similarity with
 previous and following sentences to maintain textual coherence. The method is geared towards, and was evaluated with, ranking functions that reward high query-term occurrence. Our modification approaches make no assumptions about the ranking function and are shown to be effective with both sparse and dense rankers. Furthermore, we demonstrate their rank promotion (improvement) merits with respect to Goren et al.'s method \cite{goren2020ranking}. }

%% file: bots.tex
\section{Document modification methods}
\label{sec:bots}

%\shared \contextualized

We present methods utilizing a large language model (LLM) that
modify a document so as to have it highly ranked in the next ranking induced
for a query by an undisclosed ranking function. We assume that past rankings for
the query and other queries can be observed. We post two additional
goals for the modification: having the resultant document of high
quality (in terms of discourse) and as ``faithful'' as possible to the original document in terms of the
content it contains \cite{goren2020ranking}.

The modification methods are based on prompting an LLM. The prompts have two parts: (i) a general shared part which
describes the task (document modification for improved ranking) with a
directive to maintain similarity to the original document (i.e.,
faithfulness); and, (ii) context-specific part, specific to a prompt, which provides information about past rankings. We assume that the LLM produces high quality content in terms of discourse. The general shared part is:

\input{Images/prompt}

The placeholder: <<median length of context document>> is the median number of words in documents provided in the context-specific part of the prompt that follows the general part.
%This directive is intended to avoid document-length bias in the ranking competitions.
%\newline
%\textit{\textbf{User prompt:}\newline
%<contextual information according to the method used>}\newline

\myparagraph{Context-specific part} The goal of this part of the
prompt is to provide the LLM with ``hints'' about the undisclosed
ranking function; specifically, past rankings. We present four
types of context, most of which are inspired by work on using LLMs to
induce ranking \cite{nogueira2020document, qin2023large, ma2023zero}.
\begin{itemize}
\item \firstmention{Pointwise}: The context includes pairs of a query and a
  document which was the highest ranked in the past for the
  query. This context is inspired by pointwise approaches to LLM ranking which provide the LLM with a query and examples of relevant and non-relevant documents (e.g., \cite{Liang+al:22a}). Including the documents most highly ranked in the past for the query at hand is conceptually reminiscent of a prevalent document modification strategy in competitive search \cite{raifer2017information,goren_driving_2021}: mimicking content in the most highly ranked documents in the past.
\item \firstmention{Pairwise}: Inspired by work on using pairwise approaches to prompt an LLM to induce ranking \cite{qin2023large}, the context
  includes queries accompanied with pairs of documents and an indication about which was ranked higher. 
\item \firstmention{Listwise}: Inspired by listwise prompting
  approaches for LLM ranking \cite{ma2023zero}, the context includes queries and ranked document lists. 
\item \firstmention{Temporal}: The context includes a query and
  versions of the same document and their ranks in past rankings induced for the query. The goal is to provide the LLM with information about how modifications of a document affected its ranking along time.
  %  Pairs of documents are considered (top-ranked documents or random documents);
  %This method selects
  %top-ranking documents (from one or multiple games), inspired by the
  %"mimicking the winner" strategy which has been shown to be optimal
  %in competitive search settings \cite{raifer2017information};
  \end{itemize}

\myparagraph{Prompt configurations}
There are numerous configurations
%($225$ all together)
of the context-specific part that we
study. These are determined by the following factors and parameters:
(i) the number of queries used; (ii) the number of examples used per
query: number of the most highest ranked documents in the past (Pointwise),
number of pairs (Pairwise), number of ranked lists (Listwise) and number of
documents with their temporal changes and ranks (Temporal); (iii) whether
the rank of the document to be modified in the last ranking induced
for the query is included; (iv) whether the query at hand is included
in the prompt, or alternatively only other queries; (v) for the
pairwise approach (Pairwise) we consider a random selection of the pair of
documents or the highest ranked document and an additional randomly
selected document; and, (vi) for the Temporal approach (Temporal), we vary the number of past ranks of the example document.  

%% file: Images/prompt.tex
{\em 
Edit the candidate document to improve its search engine ranking for the candidate query, aiming for the highest rank (1 being the highest). Use the black box search engine's past rankings over various queries, provided as context by the user, to guide your edits. Focus on editing the most impactful sentences to enhance ranking potential. Target an edited document length of around }<<median length of context document corpus>>\textit{ words, not exceeding 150 words. Ensure the edited document is very similar to the candidate document. Generate only the edited document, without additional comments or titles.}\newline
Input:\newline
- Candidate Query: <query>\newline
- Candidate Document: <current document>\newline
%\textit{\textbf{\contextualized:}\newline
%{\bf context-specific part:}
%<contextual information according to the method used>}\newline

%% file: experimental-setup.tex
\section{Experimental settings}
\label{sec:experimental-setup}
%We use two types of evaluation: offline and online. For offline
%evaluation,
\myparagraph{Datasets}
We use datasets which are reports of ranking
competitions \cite{raifer2017information,Mordo+al:25a}. In these
competitions, students were assigned to queries and had to produce
documents that would be highly ranked. Before the first round the students were provided with an example of a document relevant to the query. In each of the following rounds, the students observed past rankings for their queries and could modify their documents to potentially improve their next round ranking.

The first dataset, \firstmention{\firstDataset}, is the result of
ranking competitions held for $31$ queries from the TREC9-TREC12 Web
tracks \cite{raifer2017information}. Five to six students competed for each query. The undisclosed ranking function
was LambdaMART \cite{burges2010lambdamart} applied with various hand-crafted features. Following
Goren et al. \cite{goren2020ranking}, whose document modification
approach, \firstmention{\sentReplace}, serves as a baseline\footnote{We found that using LambdaMART instead of SVMrank as originally proposed \cite{goren2020ranking} yields improved performance.}, we use
round 7 for evaluation \cite{raifer2017information}. \sentReplace is a state-of-the-art feature-based supervised method for ranking-incentivized document modification. It
replaces a sentence in the document with another sentence to
improve ranking and to maintain content quality and faithfulness to
the original document.

The second dataset, \firstmention{\secondDataset}, is a report of
ranking competitions \cite{Mordo+al:25a} where the undisclosed ranking
function was the cosine between the E5 embedding vectors \cite{Wang+al:24a} of a document
and a query\footnote{The intfloat/e5-large-unsupervised version from
  the Hugging Face repository
  (\url{https://huggingface.co/intfloat/e5-large-unsupervised}).}. The competitions were run for 7 rounds with $15$
queries from the Web tracks of TREC9-TREC12; 4 players were competing
for each query \cite{wang2022text}.
%Our best performing document modification
%strategies (prompts) were used as bots in some rounds of these
%competitions for online evalution. (See more details below.)
%For
%offline evaluation,
We used round $4$ for evaluation to allow the document
modification methods to have enough history of past rankings. 

For both datasets just described, we apply the different document
modification methods, henceforth referred to as \firstmention{bots},
upon each of the documents in the ranked list for a query in the
specified round (except for the highest ranked document). For each
selected document, we induce a ranking using the same ranker used in
the competitions over its modified version and the original next-round
versions of the other documents (of students) from the round. We use
the evaluation measures described below upon the resultant ranking. We
average the evaluation results across all documents we modified per
round and over queries.

\myparagraph{Evaluation measures}
%We analyzed the performance of a document modification method using various evaluation measures, categorized into three primary groups: ranking properties, faithfulness properties and Quality and Relevance properties. All measures were computed per player and her document for a given query. The results were averaged over queries and grouped by the player type (student, baseline\footnote{i.e. the method of replacing paragraphs \cite{goren2020ranking}.}, a static document or one of the \bt s). For the online evaluation, the measures were also averaged over rounds.
To evaluate rank promotion (demotion) of documents as a result of
modification, we follow Goren et al. \cite{goren2020ranking} and
report \firstmention{Scaled Promotion}: the increase (decrease) of rank in the next round with respect to the current round normalized by the maximum possible rank promotion (demotion).

\omt{
%\begin{block}{Candidate Faithfulness at 1}
$CF@1(d_{curr},d_{next})=\frac{1}{n} \cdot \Sigma_{i=1}^{n} \mathbf{1}{\{ TT(d{curr},d_{next_{i}}) \geq 0.5 \}}$
%\end{block}

%\begin{block}{Normalized Candidate Faithfulness at 1}
$NCF@1(d_{curr},d_{next})=\frac{CF@1(d_{curr},d_{next})}{CF@1(d_{curr},d_{curr})}$
%\end{block}

%\begin{block}{Environmental Faithfulness at 10}
$EF@10(d_{next})=\frac{1}{2 \cdot 10} \cdot \Sigma_{i=1}^{10} [\mathbf{1}{\{ TT(d{\text{top}i}, d{\text{next}}) \geq 0.5 \}} + \mathbf{1}{\{ TT(d{\text{next}}, d_{\text{top}_i}) \geq 0.5 \}}]$
%\end{block}

%\begin{block}{Normalized Environmental Faithfulness at 10}
$NEF@10(d_{curr},d_{next})=\frac{EF@10(d_{next})}{EF@10(d_{curr})}$
%\end{block}
}

To evaluate the faithfulness of a modified document ($\dn$) to its
original (current) version ($\dc$), we compare
the two documents using Gekhman's et al. \cite{gekhman2023trueteacher}
natural language inference (NLI) approach. Specifically, we estimate
whether one document (denoted {\em hypothesis}) is entailed from the other
document (denoted {\em premise}) while preserving factual consistency. The estimate is the \trueteacher{} (TrueTeacher)
measure: \trueteacher $(premise, hypothesis)$ is the
output of the model in the range [0,1]; higher scores indicate stronger factual alignment.

To apply the TrueTeacher model, we first compute the average number of sentences in the modified document that are entailed\footnote{Entailment is determined by a threshold of $0.5$ for the TT score \cite{gekhman2023trueteacher}.} by the current document, which we refer to as raw faithfulness (RF):
%\trueteacher{} score between
%the current document ($\dc$) and all ($n$) sentences in the modified
%document ($\dn$):
$RF(\dn,\dc) \definedas \frac{1}{n} \sum_{i=1}^n \delta[\trueteacher
  (\dc, \dni) \ge 0.5];$ $d^{i}$ is the i'th sentence in document $d$;
$\delta$ is Kronecker's indicator function. Since $RF(\dc,\dc)$ is not
necessarily $1$, we normalize the raw
faithfulness to yield our \firstmention{\normFaith} measure: $\frac
{RF(\dn,\dc)}{RF(\dc,\dc)}$. 

Using LLMs to modify documents raises a concern
about hallucinations \cite{shuster2021retrieval}. We hence measure the
extent to which the content in the modified document is ``faithful''
to that in the entire corpus\footnote{For a corpus we use all the
  documents in all rounds prior to the round on which evaluation is
  performed.}. To that end, we treat the current document as a query,
and retrieve the top-$k$\footnote{We set $k=10$ in our experiments.}
documents in the corpus; $\topRet$ denotes the retrieved set. Retrieval is based on using cosine to compare a query
embedding and the document embedding. We use two types of embeddings:
E5 \cite{Wang+al:24a} and TF.IDF.  We define raw corpus faithfulness
(RCF) as: $RCF (\dn) \definedas \frac {1}{2k} \sum_{d \in \topRet}
(RF(\dn,d) + RF (d,\dn))$. The normalized corpus faithfulness
measure we use is: $CF (\dn) \definedas
\frac{RCF(\dn)}{RCF(\dc)}$. Using the E5 and TF.IDF embeddings results
in the \firstmention{\normCorpFaithE} and
\firstmention{\normCorpFaithT} normalized corpus faithfulness
measures, respectively.

Statistically significant differences are determined using the two-tailed paired permutation test
  with 100,000 random permutations and $p < 0.05$.

\myparagraph{Instantiating bots} For LLM we use Chat-GPT 4o
\cite{achiam2023gpt}. As described in Section
\ref{sec:bots}, there are a few parameters affecting the
instantiation of specific prompts. The number of queries is set to a
value in $\{1, 2\}$.  The number of examples per query is selected
from $\{1, 2, 3\}$. The number of past ranks (i.e., rounds) in the
Temporal prompt is selected from $\{2,3\}$. Using these
parameter values, and the other binary decision factors that affect
instantiation (see Section \ref{sec:bots}), results in $192$
different bots (prompts). In addition, we set the LLM's temperature parameter which controls potential drift to values in $\{0, 0.5, 1, 1.5, 2\}$ \cite{peeperkorn_is_2024}. 

\myparagraph{Rank promotion performance of bots} In terms of Scaled
Promotion, we found\footnote{Actual numbers are omitted due to space
  considerations and as they convery no additional insight.} that the
Pairwise bots (with random selection of document pairs) and the
Listwise bots were the best performing for both the \firstDataset and
\secondDataset datasets; the same specific instantiation of each of these two bots was
always among the top-3 performing bots for both datasets. This finding attests to the
rank-promotion effectiveness of these types of bots (prompts) for
different rankers (LambdaMART and E5). The Temporal bots (prompts), which provide rank-changes information along rounds, were less
effective (in terms of Scaled Promotion) than the Pairwise and
Listwise bots, but were more effective than the Pointwise bots. 

In what follows, we present the evaluation of the two
bots which posted for both datasets Scaled Promotion among the best three:\footnote{These bots were also the best performing in the online evaluation presented below.} 
%For efficiency considerations, we use LLama-2 with $13$B parameters
%\cite{touvron2023llama} to select the best performing
%configurations. The selection is performed with the \firstDataset
%dataset based on the scaled promotion evaluation measure. The best
%performing configurations for which we will report performance over the evaluation datasets are:

\begin{itemize}
\item Pairwise, where only the given query is included, one
  random pair of documents for each of the three last rounds is
  provided as examples, the current rank of the document is not
  used, and the temperature is set to $0.5$.
\item Listwise, where only the given query is included, two previous rounds are used, the current rank is not used, and the temperature is set to $0$.
\end{itemize}
Appendix \ref{appendix_prompt} provides the prompts for these bots.
%The fact that the pairwise and listwise approaches are the most
%effective is conceptually consistent of findings in work on using LLMs
%to induce ranking where the merits of pairwise and listwise approaches
%have been demonstrated \cite{ma2023zero,qin2023large}. For evaluation
%over \firstDataset and \secondDataset we use

\omt{
%The goal of the first phase is to identify a representative prompt for each class of prompts--that is, the prompt whose resultant agents maximize a metric related to ranking promotion. We conduct a comprehensive grid search over the 225 configurations described in Section \ref{sec:bots}, using Dataset 1. For this phase, we utilized Llama-2 with 13B parameters due to its availability \cite{touvron2023llama}. From each configuration, we constructed five agents, each with a different temperature setting for the probability model of the LLM\footnote{All other parameters of the LLM were fixed.}. The temperatures used were \{0, 0.5, 1, 1.5, 2\} and were selected based on the work of Peeperkorn et al \cite{peeperkorn_is_2024}. The selected agents compete for round 7, as was the case in Goren et al. \cite{goren2020ranking}. We do not report the detailed results of this phase due to space limitations in the paper.
}

\myparagraph{Online evaluation} The evaluation performed over the
\firstDataset and \secondDataset datasets is offline and therefore
spans a single round: the students who competed in the competition did
not respond to rankings induced over the documents we modify here. We
therefore also performed online evaluation where our instantiated
prompts competed as bots against students. We organized a ranking
competition\footnote{The competition was approved by institution and international ethics committees.} similar to that of Mordo et al. \cite{Mordo+al:25a} using 15
queries from TREC9-TREC12\footnote{These are different queries than
  those used in the \secondDataset dataset: 21, 55, 61, 64, 74, 75, 83, 96, 124,
  144, 161, 164, 166, 170, 194.}. In contrast to Mordo et al.'s
competitions \cite{Mordo+al:25a}, each game included 5 players: two-three
students, one of the two bots discussed above (Pairwise or
Listwise), and one or two static documents were created using a procedure similar to the one in Raifer et al. \cite{raifer2017information}: first, we used the query in the English Wikipedia search engine and selected a highly ranked page. We then extracted a candidate paragraph from this page, with a length of up to 150 words. Three annotators assessed the relevance of the passages, and we repeated the extraction process for each query until at least two annotators judged a paragraph as relevant. The selected paragraph was then used as a static document for the query for all students.

The students were not aware that they were competing
against bots. We applied our bots in rounds 5\footnote{Due to
  technical issues, we could not run the bots at round 4 as in the offline evaluation.}, 6 and 7 and report the
average performance over these three rounds.

We had documents in the online
evaluation judged for relevance and quality using crowdsourcing
annotators on the Connect platform of CloudResearch
\cite{noauthor_introducing_2024}. Following past work on ranking
competitions \cite{raifer2017information,goren2020ranking}, a
document's quality grade is set to $1$ if at least three out of five
English-speaking annotators marked it as valid (as opposed to keyword
stuffed or useless) and to $0$ otherwise. The relevance grade was $1$ if the document was
marked relevant by at least three annotators and $0$ otherwise.

%% file: experimental-results.tex
\section{Experimental Results}
\label{sec:experimental-results}

\begin{table}[t]
% \centering
  \caption{\label{table_offline2017} Evaluation over the \firstDataset
    dataset which resulted from competitions using a LambdaMART
    ranker \cite{raifer2017information}. The baseline (\sentReplace \cite{goren2020ranking}), Pairwise and Listwise blocks correspond to a
    round where a document was modified by the respective bot, and all other documents were of
    students. Boldface marks the best result in a column.
    Statistically significant differences of our bots with \sentReplace and the students are marked with '$\baseDiff$' and '$\studDiff$', respectively.}
    \scriptsize
%\hspace{-15mm}
%\input{Tables/offline2017}
\input{Tables/offline2017,v2}
\end{table}

Table \ref{table_offline2017} presents the evaluation over the
\firstDataset. We let one of our bots, or the baseline
bot, \sentReplace \cite{goren2020ranking}, modify a single document at a time. We then
analyze the resultant ranking over the single modified document and
the next-round students' documents from the dataset. The performance numbers are averages over the different selected documents to be modified (all but the highest ranked one) and over queries. Note that the Students' faithfulness values are not block dependent as they are the same in all three cases in the next round. Their Scaled Promotion is affected by the bot used to modify the selected document.

Table \ref{table_offline2017} shows that our Pairwise and Listwise
bots outperform \sentReplace and the students in terms of Scaled
Promotion. The faithfulness of both of our bots is lower than that of
the students and \sentReplace but is still quite high. Note that \sentReplace modifies a single sentence
in a document and, hence, the modified document is quite faithful to the original
document. The corpus faithfulness of both our bots (for both TF.IDF and E5) is higher than that of \sentReplace. Thus, although our bots use an LLM (as opposed to \sentReplace and the students), their corpus faithfulness is relatively high.

Table \ref{table_offline2024} presents the evaluation for the
\secondDataset dataset\footnote{We do not use here and after the
  \sentReplace baseline as it was defined for sparse ranking
  functions.}. The results presented above for \firstDataset show that the bots were not better than the students in terms of corpus faithfulness. Over the \secondDataset, our bots outperform the students with respect to
Scaled Promotion and corpus faithfulness and underperform them in
terms of faithfulness to the original document, although this faithfulness is still relatively high. All in all, Tables
\ref{table_offline2017} and \ref{table_offline2024} attest to the
effectiveness of our bots for both ranking functions: LambdaMART and
E5.

Table \ref{table_online2024} presents the evaluation for the
online competitions (our bots vs. students) which used the E5 ranker.
% We use for reference a
% Static bot which uses the example document provided to students at the
% beginning of the competition and does not modify it along
% rounds; hence, its \normFaith is $1$.
Q and R are the average
quality and relevance grades, respectively, over documents, queries and the three
rounds ($5$-$7$). We see in Table \ref{table_online2024} that our bots
outperform the students in terms of Scaled Promotion, \normFaith and
corpus faithfulness (except for a single case). The documents produced by our Listwise bots are of higher
quality on average than those produced by students in the same
competitions. For the competitions with our Pairwise bot, the
students produced documents of higher quality on average than those of the bot; yet, the bot
 produced in a majority of cases ($0.822$) quality documents. Finally, our bots produce
documents that are relevant in more cases than those produced by the
students.

All in all, the findings presented above attest to the clear merits of our
bots with respect to students and the \sentReplace baseline along various dimensions; specifically, rank promotion, document quality and relevance, and faithfulness to the corpus.

\begin{table}[t]
% \centering
\caption{\label{table_offline2024} Evaluation over the \secondDataset dataset which resulted from
  competitions using an E5 ranker. The Pairwise and Listwise blocks correspond to a round where a document was modified by the respective bot and all other documents were of students.
    Boldface: the best result in a column.
    Statistically significant differences of our bots with the students are marked with '$\studDiff$'.}
%  The Pairwise and Listwise blocks correspond to a round where a document was modified by the Pairwise and Listwise approach, respectively, and all other documents were modified by students.
%    Boldface marks the best result in a column.
%    Statistically significant differences of our bots with the students and the baseline are marked with '$\studDiff$' and '$\baseDiff$', respectively.}
\scriptsize
%\hspace{-15mm}
%\input{Tables/offline2024}
\input{Tables/offline2024,v2}
\end{table}

\setlength{\tabcolsep}{2pt}
\begin{table}[t]
  % \centering
  \caption{\label{table_online2024} Online evaluation. Q and R are the
    average quality and relevance grades, respectively. The Pairwise
    and Listwise blocks correspond to competitions where one of the
    two bots participated. Boldface marks the best result in a
    column. Statistically significant differences of our bots with the students are marked with '$\studDiff$'.}
  \scriptsize
\input{Tables/online2024,v2}

\end{table}

%% file: Tables/offline2017,v2.tex
\begin{tabular}{lcccc}
\toprule
             & {Scaled Promotion} & \normFaith & \corpFaithE & \corpFaithT \\
\midrule
\multicolumn{5}{c}{\sentReplace \cite{goren2020ranking}} \\ \midrule
\sentReplace  & $0.309$ & $\mathbf{0.786}$ & $0.468$ & $0.381$\\
Students & $0.313$ & $0.727$ &   $\mathbf{0.544}$ & $\mathbf{0.504}$ \\ \midrule
\multicolumn{5}{c}{Pairwise} \\ \midrule
       Our bot & $\mathbf{0.345}^{\studDiff}$ & $0.57_{\baseDiff}^{\studDiff}$ &   $0.507$ & $0.437$ \\
       Students & $-0.095$ & $0.727$ &  $\mathbf{0.544}$ & $\mathbf{0.504}$ \\ \midrule
       \multicolumn{5}{c}{Listwise} \\ \midrule
       Our bot & $0.315^{\studDiff}$ & $0.616^{\studDiff}_{\baseDiff}$   & $0.497$ & $0.427$ \\
       Students & $-0.09$ & $0.727$ & $\mathbf{0.544}$ & $\mathbf{0.504}$\\
\bottomrule
\end{tabular}

% \begin{tabular}{lrrrrrr}
% \toprule
%             tag & \makecell{average \\ rank} & \makecell{raw \\ promotion}  & \makecell{scaled \\ promotion} & NCF@1 & \makecell{NEF@10 \\ dense} & \makecell{NEF@10 \\ sparse} \\
% \midrule
%        baseline                  & 2.669 & 0.831 & 0.309 & 0.786 & 0.381 & 0.468 \\
%        pairwise random           & 2.516^{s} & 0.984^{s} & 0.345^{s} & 0.57^{s,b} & 0.437 & 0.507 \\
%        student - pairwise random & 3.144 & -0.244 & -0.095 & 0.727& 0.504 & 0.544 \\
%        listwise                  & 2.597^{s} & 0.903^{s} & 0.315^{s} & 0.616^{s,b} & 0.427 & 0.497 \\
%        student - listwise        & 3.124 & -0.224 & -0.09 & 0.727 & 0.504 & 0.544 \\
% \bottomrule
% \end{tabular}

%% file: Tables/offline2024,v2.tex
\begin{tabular}{lcccc}
\toprule
             & {Scaled Promotion} & \normFaith & \normCorpFaithE & \normCorpFaithT \\
\midrule
\multicolumn{5}{c}{Pairwise} \\\midrule
       Our bot    & ${0.241}^{\studDiff}$  &  ${0.623}^{\studDiff}$  &  $\mathbf{0.589}$  &  $\mathbf{0.603}$  \\
       Students & $-0.107$  &$  \mathbf{0.788}$  &  $0.531$  &  $0.542$  \\ \midrule
       \multicolumn{5}{c}{Listwise}  \\ \midrule
       Our bot    & $\mathbf{0.341}^{\studDiff}$  &  $0.711$  &  $0.573$  &  $0.592$  \\
       Students    & $-0.128$  &  $\mathbf{0.788}$  &  $0.531$  &  $0.542$  \\
\bottomrule
\end{tabular}

% \begin{tabular}{lrrrrrr}
% \toprule
%             tag & \makecell{average \\ rank} & \makecell{raw \\ promotion}  & \makecell{scaled \\ promotion} & NCF@1 & \makecell{NEF@10 \\ dense} & \makecell{NEF@10 \\ sparse} \\
% \midrule
%        pairwise random           & 2.533  &  0.467^{s}  &  0.241^{s}  &  0.623  &  0.589  &  0.603  \\
%        student - pairwise random & 2.481  & -0.148  & -0.107  &  0.788  &  0.531  &  0.542  \\
%        listwise                  & 2.378  &  0.622^{s}  &  0.341^{s}  &  0.711  &  0.573  &  0.592  \\
%        student - listwise        & 2.533  & -0.200  & -0.128  &  0.788  &  0.531  &  0.542  \\
% \bottomrule
% \end{tabular}

%% file: Tables/online2024,v2.tex
\begin{tabular}{lcccccc}
\toprule & {Scaled Promotion} & \normFaith & \corpFaithE & \corpFaithT
& Q & R \\ \midrule \multicolumn{6}{c}{Pairwise} \\ \midrule Our bot &
$\mathbf{0.098}_{\statDif}$ & ${0.839}_{\statDif}$ &
$0.663_{\statDif}$ & $0.735_{\statDif}$ & $0.822$ & $\mathbf{0.8}_{\statDif}$
\\ Students & $0.029$ & $0.821$ & $0.675$ & $0.726$ & $\mathbf{0.925}$ &
$0.642$
% \\ Static bot & $-0.069$ & $\mathbf{1.0}$ & $0.538$ & $0.554$ & $0.842$
% & $0.316$
\\ \midrule \multicolumn{6}{c}{Listwise} \\ \midrule Our bot
& $0.05$ & $\mathbf{0.875}_{\statDif}$ & $\mathbf{0.717}_{\statDif}$ &
$\mathbf{0.756}_{\statDif}$ & ${0.911}^{\studDiff}$ &
$\mathbf{0.8}_{\statDif}$ \\ Students & $-0.013$ & $0.865$ & $0.685$ & $0.754$
& $0.732$ & $0.66$
% \\ Static bot & $-0.049$ & $\mathbf{1.0}$ & $0.533$ &
% $0.548$ & $0.9$ & $0.35$
\\ \bottomrule
\end{tabular}

%% file: conclusions.tex
\section{Conclusions}
\label{sec:conclusions}
We presented novel methods for document
modifications: modifying a document so it will be highly ranked by an undisclosed ranking function for a query. Our methods are based on prompting large language models (LLMs). Some of our methods are inspired by prompting approaches for inducing LLM-based ranking over documents.

We conducted extensive empirical evaluation using past ranking
competitions (for two different rankers). In addition, we organized ranking competitions where our document modification methods competed as bots against students.

The empirical evaluation demonstrated the merits of our best
performing modification methods with respect to students and a
previously proposed highly effective feature-based document
modification approach.

%% file: appendix.tex
\appendix
\section{Prompts}\label{appendix_prompt}

\begin{figure}[H]
\centering
\input{Images/pairwise}
\caption{Context-specific part of the Pairwise prompt which includes document pairs from the last three rankings. For each ranking, two documents, (a,b), (c,d), (e,f), were randomly selected and their rank is specified: r(a), r(b), r(c), r(d), r(e), r(f).}
\label{prompt_pairwise}
\end{figure}

\begin{figure}[H]
\centering
\input{Images/listwise}
\caption{Context-specific part of the Listwise prompt which includes the latest ranking over documents: a, b, c, d (excluding the current document) and the previous ranking of the entire document list: e, f, g, h, i.}
\label{prompt_listwise}
\end{figure}

%\endinput

%% file: Images/pairwise.tex
% \begin{minted}[breaklines=true, frame=single, fontsize=\tiny]{python}
% PAIRWISE_PROMPT = "\n\nquery: <QUERY>\n\n* document: <DOCUMENT a>\n\nlatest ranking: <RANK r(a)>\n\n\n* document: <DOCUMENT b>\n\nlatest ranking: <RANK r(b)>\n\n\n\n\nquery: <QUERY>\n\n* document: <DOCUMENT c>\n\nsecond to latest ranking: <RANK r(c)>\n\n\n* document: <DOCUMENT d>\n\nsecond to latest ranking: <RANK r(d)>\n\n\n\n\nquery: <QUERY>\n\n* document: <DOCUMENT e>\n\nthird to latest ranking: <RANK r(e)>\n\n\n* document: <DOCUMENT f>\n\nthird to latest ranking: <RANK r(f)>"
% \end{minted}

\begin{lstlisting}[style=promptstyle]
PAIRWISE_PROMPT = "\n\nquery: <QUERY>\n\n* document: <DOCUMENT a>\n\nlatest ranking: <RANK r(a)>\n\n\n* document: <DOCUMENT b>\n\nlatest ranking: <RANK r(b)>\n\n\n\n\nquery: <QUERY>\n\n* document: <DOCUMENT c>\n\nsecond to latest ranking: <RANK r(c)>\n\n\n* document: <DOCUMENT d>\n\nsecond to latest ranking: <RANK r(d)>\n\n\n\n\nquery: <QUERY>\n\n* document: <DOCUMENT e>\n\nthird to latest ranking: <RANK r(e)>\n\n\n* document: <DOCUMENT f>\n\nthird to latest ranking: <RANK r(f)>"
\end{lstlisting}

%% file: Images/listwise.tex
% \begin{minted}[breaklines=true, frame=single, fontsize=\tiny]{python}
% LISTWISE_PROMPT = "\n\nquery: <QUERY>\n\n* documents ordered by latest ranking from highest to lowest in relation to the query: \n\n\n* <DOCUMENT a>\n\n\n* <DOCUMNET b>\n\n\n* <DOCUMENT c>\n\n\n* <DOCUMENT d>\n\n\n\n\n\n * documents ranked by second to latest ranking from highest to lowest in relation to the query:\n1. <DOCUMENT e>\n2. <DOCUMENT f>\n3. <DOCUMENT g>\n4. <DOCUMENT h>\n5. <DOCUMENT i>\n\n\n\n"
% \end{minted}

\begin{lstlisting}[style=promptstyle]
LISTWISE_PROMPT = "\n\nquery: <QUERY>\n\n* documents ordered by latest ranking from highest to lowest in relation to the query: \n\n\n* <DOCUMENT a>\n\n\n* <DOCUMNET b>\n\n\n* <DOCUMENT c>\n\n\n* <DOCUMENT d>\n\n\n\n\n\n * documents ranked by second to latest ranking from highest to lowest in relation to the query:\n1. <DOCUMENT e>\n2. <DOCUMENT f>\n3. <DOCUMENT g>\n4. <DOCUMENT h>\n5. <DOCUMENT i>\n\n\n\n"
\end{lstlisting}

%% file: main.bbl

%% file: main.bbl
\begin{thebibliography}{23}

%%% the Web address.

\ifx \showCODEN    \undefined \def \showCODEN     #1{\unskip}     \fi
\ifx \showDOI      \undefined \def \showDOI       #1{#1}\fi
\ifx \showISBNx    \undefined \def \showISBNx     #1{\unskip}     \fi
\ifx \showISBNxiii \undefined \def \showISBNxiii  #1{\unskip}     \fi
\ifx \showISSN     \undefined \def \showISSN      #1{\unskip}     \fi
\ifx \showLCCN     \undefined \def \showLCCN      #1{\unskip}     \fi
\ifx \shownote     \undefined \def \shownote      #1{#1}          \fi
\ifx \showarticletitle \undefined \def \showarticletitle #1{#1}   \fi
\ifx \showURL      \undefined \def \showURL       {\relax}        \fi
% invisible to TeX
\providecommand\bibfield[2]{#2}
\providecommand\bibinfo[2]{#2}
\providecommand\natexlab[1]{#1}
\providecommand\showeprint[2][]{arXiv:#2}

\bibitem[noa(2024)]%
        {noauthor_introducing_2024}
 \bibinfo{year}{2024}\natexlab{}.
\newblock \bibinfo{title}{Introducing {Connect} by {CloudResearch}: {Advancing} {Online} {Participant} {Recruitment} in the {Digital} {Age} {\textbar} {Request} {PDF}}.
\newblock
\newblock
\urldef\tempurl%
\url{https://doi.org/10.31234/osf.io/ksgyr}
\showDOI{\tempurl}


\bibitem[Achiam et~al\mbox{.}(2023)]%
        {achiam2023gpt}
\bibfield{author}{\bibinfo{person}{Josh Achiam}, \bibinfo{person}{Steven Adler}, \bibinfo{person}{Sandhini Agarwal}, \bibinfo{person}{Lama Ahmad}, \bibinfo{person}{Ilge Akkaya}, \bibinfo{person}{Florencia~Leoni Aleman}, \bibinfo{person}{Diogo Almeida}, \bibinfo{person}{Janko Altenschmidt}, \bibinfo{person}{Sam Altman}, \bibinfo{person}{Shyamal Anadkat}, {et~al\mbox{.}}} \bibinfo{year}{2023}\natexlab{}.
\newblock \showarticletitle{Gpt-4 technical report}.
\newblock \bibinfo{journal}{\emph{arXiv preprint arXiv:2303.08774}} (\bibinfo{year}{2023}).
\newblock


\bibitem[Aggarwal et~al\mbox{.}(2024)]%
        {Aggarwal+al:24a}
\bibfield{author}{\bibinfo{person}{Pranjal Aggarwal}, \bibinfo{person}{Vishvak Murahari}, \bibinfo{person}{Tanmay Rajpurohit}, \bibinfo{person}{Ashwin Kalyan}, \bibinfo{person}{Karthik Narasimhan}, {and} \bibinfo{person}{Ameet Deshpande}.} \bibinfo{year}{2024}\natexlab{}.
\newblock \showarticletitle{{GEO:} Generative Engine Optimization}. In \bibinfo{booktitle}{\emph{Proceedings of KDD}}. \bibinfo{publisher}{{ACM}}, \bibinfo{pages}{5--16}.
\newblock


\bibitem[Burges(2010)]%
        {burges2010lambdamart}
\bibfield{author}{\bibinfo{person}{Christopher~JC Burges}.} \bibinfo{year}{2010}\natexlab{}.
\newblock \showarticletitle{From ranknet to lambdarank to lambdamart: An overview}.
\newblock \bibinfo{journal}{\emph{Learning}} \bibinfo{volume}{11}, \bibinfo{number}{23-581} (\bibinfo{year}{2010}), \bibinfo{pages}{81}.
\newblock


\bibitem[Gekhman et~al\mbox{.}(2023)]%
        {gekhman2023trueteacher}
\bibfield{author}{\bibinfo{person}{Zorik Gekhman}, \bibinfo{person}{Jonathan Herzig}, \bibinfo{person}{Roee Aharoni}, \bibinfo{person}{Chen Elkind}, {and} \bibinfo{person}{Idan Szpektor}.} \bibinfo{year}{2023}\natexlab{}.
\newblock \bibinfo{title}{TrueTeacher: Learning Factual Consistency Evaluation with Large Language Models}.
\newblock
\newblock
\showeprint[arxiv]{2305.11171}~[cs.CL]


\bibitem[Goren et~al\mbox{.}(2020)]%
        {goren2020ranking}
\bibfield{author}{\bibinfo{person}{Gregory Goren}, \bibinfo{person}{Oren Kurland}, \bibinfo{person}{Moshe Tennenholtz}, {and} \bibinfo{person}{Fiana Raiber}.} \bibinfo{year}{2020}\natexlab{}.
\newblock \showarticletitle{Ranking-incentivized quality preserving content modification}. In \bibinfo{booktitle}{\emph{Proceedings of SIGIR}}. \bibinfo{pages}{259--268}.
\newblock


\bibitem[Goren et~al\mbox{.}(2021)]%
        {goren_driving_2021}
\bibfield{author}{\bibinfo{person}{Gregory Goren}, \bibinfo{person}{Oren Kurland}, \bibinfo{person}{Moshe Tennenholtz}, {and} \bibinfo{person}{Fiana Raiber}.} \bibinfo{year}{2021}\natexlab{}.
\newblock \showarticletitle{Driving the {Herd}: {Search} {Engines} as {Content} {Influencers}}. In \bibinfo{booktitle}{\emph{Proceedings of CIKM}}. \bibinfo{address}{Virtual Event Queensland Australia}, \bibinfo{pages}{586--595}.
\newblock


\bibitem[Gy{\"o}ngyi and Garcia-Molina(2005)]%
        {Gyongyi+Molina:05a}
\bibfield{author}{\bibinfo{person}{Zolt{\'a}n Gy{\"o}ngyi} {and} \bibinfo{person}{Hector Garcia-Molina}.} \bibinfo{year}{2005}\natexlab{}.
\newblock \showarticletitle{Web Spam Taxonomy}. In \bibinfo{booktitle}{\emph{Proceedings of AIRWeb 2005}}. \bibinfo{pages}{39--47}.
\newblock


\bibitem[Kumar and Lakkaraju(2024)]%
        {Kumar+Lakkaraju:24a}
\bibfield{author}{\bibinfo{person}{Aounon Kumar} {and} \bibinfo{person}{Himabindu Lakkaraju}.} \bibinfo{year}{2024}\natexlab{}.
\newblock \showarticletitle{Manipulating Large Language Models to Increase Product Visibility}.
\newblock \bibinfo{journal}{\emph{CoRR}}  \bibinfo{volume}{abs/2404.07981} (\bibinfo{year}{2024}).
\newblock


\bibitem[Kurland and Tennenholtz(2022)]%
        {Kurland+Tennnholtz:22a}
\bibfield{author}{\bibinfo{person}{Oren Kurland} {and} \bibinfo{person}{Moshe Tennenholtz}.} \bibinfo{year}{2022}\natexlab{}.
\newblock \showarticletitle{Competitive {Search}}. In \bibinfo{booktitle}{\emph{Proceedings of SIGIR}}. \bibinfo{pages}{2838--2849}.
\newblock


\bibitem[Liang et~al\mbox{.}(2022)]%
        {Liang+al:22a}
\bibfield{author}{\bibinfo{person}{Percy Liang}, \bibinfo{person}{Rishi Bommasani}, \bibinfo{person}{Tony Lee}, \bibinfo{person}{Dimitris Tsipras}, \bibinfo{person}{Dilara Soylu}, \bibinfo{person}{Michihiro Yasunaga}, \bibinfo{person}{Yian Zhang}, \bibinfo{person}{Deepak Narayanan}, \bibinfo{person}{Yuhuai Wu}, {and} \bibinfo{person}{Ananya Kumar}.} \bibinfo{year}{2022}\natexlab{}.
\newblock \bibinfo{title}{Holistic evaluation of language models}.
\newblock
\newblock
\showeprint[arxiv]{arXiv:2211.09110}


\bibitem[Ma et~al\mbox{.}(2023)]%
        {ma2023zero}
\bibfield{author}{\bibinfo{person}{Xueguang Ma}, \bibinfo{person}{Xinyu Zhang}, \bibinfo{person}{Ronak Pradeep}, {and} \bibinfo{person}{Jimmy Lin}.} \bibinfo{year}{2023}\natexlab{}.
\newblock \showarticletitle{Zero-shot listwise document reranking with a large language model}.
\newblock \bibinfo{journal}{\emph{arXiv preprint arXiv:2305.02156}} (\bibinfo{year}{2023}).
\newblock


\bibitem[Mordo et~al\mbox{.}(2025)]%
        {Mordo+al:25a}
\bibfield{author}{\bibinfo{person}{Tommy Mordo}, \bibinfo{person}{Itamar Reinman}, \bibinfo{person}{Moshe Tennenholtz}, {and} \bibinfo{person}{Oren Kurland}.} \bibinfo{year}{2025}\natexlab{}.
\newblock \bibinfo{title}{Search results diversification in competitive search}.
\newblock
\newblock
\showeprint[arxiv]{2501.14922}


\bibitem[Nestaas et~al\mbox{.}(2024)]%
        {Nestaas+al:24a}
\bibfield{author}{\bibinfo{person}{Fredrik Nestaas}, \bibinfo{person}{Edoardo Debenedetti}, {and} \bibinfo{person}{Florian Tram{\`{e}}r}.} \bibinfo{year}{2024}\natexlab{}.
\newblock \showarticletitle{Adversarial Search Engine Optimization for Large Language Models}.
\newblock \bibinfo{journal}{\emph{CoRR}}  \bibinfo{volume}{abs/2406.18382} (\bibinfo{year}{2024}).
\newblock


\bibitem[Nogueira et~al\mbox{.}(2020)]%
        {nogueira2020document}
\bibfield{author}{\bibinfo{person}{Rodrigo Nogueira}, \bibinfo{person}{Zhiying Jiang}, {and} \bibinfo{person}{Jimmy Lin}.} \bibinfo{year}{2020}\natexlab{}.
\newblock \showarticletitle{Document ranking with a pretrained sequence-to-sequence model}.
\newblock \bibinfo{journal}{\emph{arXiv preprint arXiv:2003.06713}} (\bibinfo{year}{2020}).
\newblock


\bibitem[Peeperkorn et~al\mbox{.}(2024)]%
        {peeperkorn_is_2024}
\bibfield{author}{\bibinfo{person}{Max Peeperkorn}, \bibinfo{person}{Tom Kouwenhoven}, \bibinfo{person}{Dan Brown}, {and} \bibinfo{person}{Anna Jordanous}.} \bibinfo{year}{2024}\natexlab{}.
\newblock \bibinfo{title}{Is {Temperature} the {Creativity} {Parameter} of {Large} {Language} {Models}?}
\newblock
\newblock
\urldef\tempurl%
\url{https://doi.org/10.48550/arXiv.2405.00492}
\showDOI{\tempurl}
\newblock
\shownote{arXiv:2405.00492 [cs]}.


\bibitem[Pfrommer et~al\mbox{.}(2024)]%
        {Pfrommer+al:24a}
\bibfield{author}{\bibinfo{person}{Samuel Pfrommer}, \bibinfo{person}{Yatong Bai}, \bibinfo{person}{Tanmay Gautam}, {and} \bibinfo{person}{Somayeh Sojoudi}.} \bibinfo{year}{2024}\natexlab{}.
\newblock \showarticletitle{Ranking Manipulation for Conversational Search Engines}. In \bibinfo{booktitle}{\emph{Proceedings of EMNLP}}. \bibinfo{pages}{9523--9552}.
\newblock


\bibitem[Qin et~al\mbox{.}(2023)]%
        {qin2023large}
\bibfield{author}{\bibinfo{person}{Zhen Qin}, \bibinfo{person}{Rolf Jagerman}, \bibinfo{person}{Kai Hui}, \bibinfo{person}{Honglei Zhuang}, \bibinfo{person}{Junru Wu}, \bibinfo{person}{Jiaming Shen}, \bibinfo{person}{Tianqi Liu}, \bibinfo{person}{Jialu Liu}, \bibinfo{person}{Donald Metzler}, \bibinfo{person}{Xuanhui Wang}, {et~al\mbox{.}}} \bibinfo{year}{2023}\natexlab{}.
\newblock \showarticletitle{Large language models are effective text rankers with pairwise ranking prompting}.
\newblock \bibinfo{journal}{\emph{arXiv preprint arXiv:2306.17563}} (\bibinfo{year}{2023}).
\newblock


\bibitem[Raifer et~al\mbox{.}(2017)]%
        {raifer2017information}
\bibfield{author}{\bibinfo{person}{Nimrod Raifer}, \bibinfo{person}{Fiana Raiber}, \bibinfo{person}{Moshe Tennenholtz}, {and} \bibinfo{person}{Oren Kurland}.} \bibinfo{year}{2017}\natexlab{}.
\newblock \showarticletitle{Information retrieval meets game theory: The ranking competition between documents' authors}. In \bibinfo{booktitle}{\emph{Proceedings of SIGIR}}. \bibinfo{pages}{465--474}.
\newblock


\bibitem[Shuster et~al\mbox{.}(2021)]%
        {shuster2021retrieval}
\bibfield{author}{\bibinfo{person}{Kurt Shuster}, \bibinfo{person}{Spencer Poff}, \bibinfo{person}{Moya Chen}, \bibinfo{person}{Douwe Kiela}, {and} \bibinfo{person}{Jason Weston}.} \bibinfo{year}{2021}\natexlab{}.
\newblock \showarticletitle{Retrieval augmentation reduces hallucination in conversation}. In \bibinfo{booktitle}{\emph{Findings of the Association for Computational Linguistics: EMNLP 2021}}. \bibinfo{pages}{3784--3803}.
\newblock


\bibitem[Wang et~al\mbox{.}(2022)]%
        {wang2022text}
\bibfield{author}{\bibinfo{person}{Liang Wang}, \bibinfo{person}{Nan Yang}, \bibinfo{person}{Xiaolong Huang}, \bibinfo{person}{Binxing Jiao}, \bibinfo{person}{Linjun Yang}, \bibinfo{person}{Daxin Jiang}, \bibinfo{person}{Rangan Majumder}, {and} \bibinfo{person}{Furu Wei}.} \bibinfo{year}{2022}\natexlab{}.
\newblock \bibinfo{title}{Text Embeddings by Weakly-Supervised Contrastive Pre-training. CoRR abs/2212.03533 (2022)}.
\newblock
\newblock


\bibitem[Wang et~al\mbox{.}(2024)]%
        {Wang+al:24a}
\bibfield{author}{\bibinfo{person}{Liang Wang}, \bibinfo{person}{Nan Yang}, \bibinfo{person}{Xiaolong Huang}, \bibinfo{person}{Binxing Jiao}, \bibinfo{person}{Linjun Yang}, \bibinfo{person}{Daxin Jiang}, \bibinfo{person}{Rangan Majumder}, {and} \bibinfo{person}{Furu Wei}.} \bibinfo{year}{2024}\natexlab{}.
\newblock \bibinfo{title}{Text Embeddings by Weakly-Supervised Contrastive Pre-training}.
\newblock
\newblock
\showeprint[arxiv]{2212.03533}


\bibitem[Wu et~al\mbox{.}(2023)]%
        {Wu+al:23a}
\bibfield{author}{\bibinfo{person}{Chen Wu}, \bibinfo{person}{Ruqing Zhang}, \bibinfo{person}{Jiafeng Guo}, \bibinfo{person}{Maarten de Rijke}, \bibinfo{person}{Yixing Fan}, {and} \bibinfo{person}{Xueqi Cheng}.} \bibinfo{year}{2023}\natexlab{}.
\newblock \showarticletitle{{PRADA:} Practical Black-box Adversarial Attacks against Neural Ranking Models}.
\newblock \bibinfo{journal}{\emph{{ACM} Trans. Inf. Syst.}} \bibinfo{volume}{41}, \bibinfo{number}{4} (\bibinfo{year}{2023}), \bibinfo{pages}{89:1--89:27}.
\newblock


\end{thebibliography}
